\documentstyle[eqsecnum,aps,psfig]{revtex}

\def\etal{{\it et\thinspace al.}\ }
\begin{document}
\draft
\preprint{HEP/123-qed}
\title{Photoionization of Fe XV}
\author{Nasreen Haque}
\address{
Department of Physics, Morehouse College, Atlanta, GA 30314\\
}
\author{Anil K. Pradhan}
\address{
Department of Astronomy, The Ohio State University, Columbus, Ohio
43210\\
}
\date{\today}
\maketitle
\begin{abstract}
 Relativistic and resonance effects in the photoionization of Mg-like Fe~XV
are investigated using the Breit-Pauli R-matrix Method (BPRM) at 
near-threshold and intermediate energies, complemented by the
Relativistic Random Phase Approximation (RRPA) and multi-channel quantum
defect theory in the energy region up to the L-shell ionization
thresholds. The cross sections exhibit extensive resonance structures that
considerably enhance the effective
photoionization of Fe~XV. These results should be of general interest in
photoionization modeling of X-ray sources observed by space
observatories.

\end{abstract}
\pacs{PACS number(s): 32.80.Fb, 32.30.Rj, 98.58.Bz, 98.70.Qy}

\narrowtext

 There is particular interest  in the photoionization of highly
charged iron ions owing to their importance in the modeling
of astrophysical
plasmas such as in active galactic nuclei and quasars, supernovae, stellar
coronae etc., and in laboratory sources 
such as inertial and magnetic confinement fusion devices [1]. Among the
most complex atomic species are those
isoelectronic with the third row elements containing open n = 3 shells. 
The K- and L-shell photoexcitations and photoionizations are highly
topical areas of investigation as they are generally in
the X-ray range accessible to satellite observations, 
such as from the recently launched Chandra X-ray
Observatory (CXO), and the upcoming X-ray Multi-Mirror Mission (XMM) 
and Astro-E [1].

While previous calculations exist for the iron ions, they are in
relatively simpler approximations that neglect the multiplet or the fine
structure, and autoionizing resonances that are known to
enhance the cross sections significantly [2]. 
Recently, extensive photoionization calculations were
carried out in the close coupling approximation using the R-matrix
method under the Opacity Project [OP,3] for most astrophysically 
abundant elements.  While the OP cross sections include the
resonance effects, they are in 
LS coupling neglecting fine structure effects
in the detailed cross sections, as
shown in this work through relativistic close coupling calculations
using the Breit-Pauli R-matrix codes developed for the Iron Project
[4].

 Whereas the low energy region can be accurately considered with the
BPRM method, at higher energies leading up to the L- and K-shell ionization
the number of coupled channels becomes very
large and the computations quite prohibitive. On the other hand, there
are very few studies done of the resonance structures leading up to
these ionization edges [5]. It is therefore of interest to 
not only carry out the
low energy BPRM calculations, but also to study the nature of resonances
at high energies using other methods such as the relativistic random
phase approximation (RRPA) and
relativistic multi-channel quantum defect theory (RMQDT). 
The aim is to combine the two techniques to investigate
photoionization cross sections over the entire energy range of practical
interest, as a guide to more elaborate close coupling calculations.

 The computationally intensive BPRM calculations are
carried out as described in recent works [6]. The residual (target) ion
following photoionization is represented by an $N$-electron system, and the 
total wavefunction of the initial bound
($N$+1) electron-ion system  is
represented in terms of the ion eigenfunctions as:

\begin{equation}
\Psi(E) = A \sum_{i} \chi_{i}\theta_{i} + \sum_{j} c_{j} \Phi_{j},
\end{equation}

\noindent
where $\chi_{i}$ is the ion wavefunction in a specific state
$S_iL_iJ_i\pi_i$ and $\theta_{i}$ is the wavefunction for the
($N$+1)-th electron in a channel labeled as
$S_iL_i(J_i)\pi_ik_{i}^{2}\ell_i \ \ [J\pi]$;
$k_{i}^{2}$
being its incident kinetic energy. The $\Phi_j$'s are the correlation
functions of the ($N$+1)-electron system that account for short range
correlation and the orthogonality between the continuum and the bound
orbitals.  The $\Phi_j$'s may also give rise to bound channel resonances
at intermediate energies due to inner-shell
photoexcitation-autoionization, as described in this work.

 The eigenfunctions ${\chi_i}$ of the residual ion Fe~XVI are
obtained through atomic structure calculations using the program
SUPERSTRUCTURE [7]. Table 1 lists the ${\chi_i}$ corresponding to 17 levels
dominated by configurations $2p^6 \ (^1S) \ n \ell$,
up to n = 5. The calculated
eigenenergies are compared with the experimetal values. Although the
Fe~XVI levels included in the close coupling expansion span the n = 5
levels, most of the relevant resonance structure appears below the first
excited fine structure levels $^2P_o{1/2,3/2}$. However, owing to the high ion
charge the n-complexes are well separated and the coupling effects,
including resonances at higher energies, are weak.

 We consider the photoionization of the ground level of Fe~XV, 
$3s^2 \ (^1S_0)  + h\nu \longrightarrow e + 3s \ (^2S_{1/2})$, coupled to
the continua of the ground and the excited levels of Fe~XVI. 
 Fig. 1 shows the detailed cross section with the Rydberg series of
resonances in the region up to, and
slightly above, the first two excited thresholds $3p \ (^2P^o_{1/2,3/2})$. 
For comparison, the non-relativistic cross sections from the Opacity
Project work by Butler \etal [8] are also shown (dashed lines). It is
seen that the OP results, in LS coupling,  did not resolve the extensive resonance structures obtained in the present calculations including
relativistic fine structure.
In the region 
$^2P^o_{1/2} - ^2P^o_{3/2}$ the resonances are considerably weaker than 
in the region
below the $^2P^o_{1/2}$ owing to autoionization into the excited $3p \ ^2P^o_{1/2}$
continuum, in addition to the ground level continuum $3s \ ^2S_{1/2}$). 
Fig. 2 shows the photoionization cross section in an extended energy
range that spans all excited Fe~XVI levels in the close coupling
expansion (Table 1). Above the $3p \ ^2P^o_{1/2,3/2}$ thresholds however,
resonance structures are fairly isolated and relatively far apart.
The background cross sections and some resonance positions
agree well with the earlier LS coupling OP results [8].

\begin{figure}
\centering
\psfig{figure=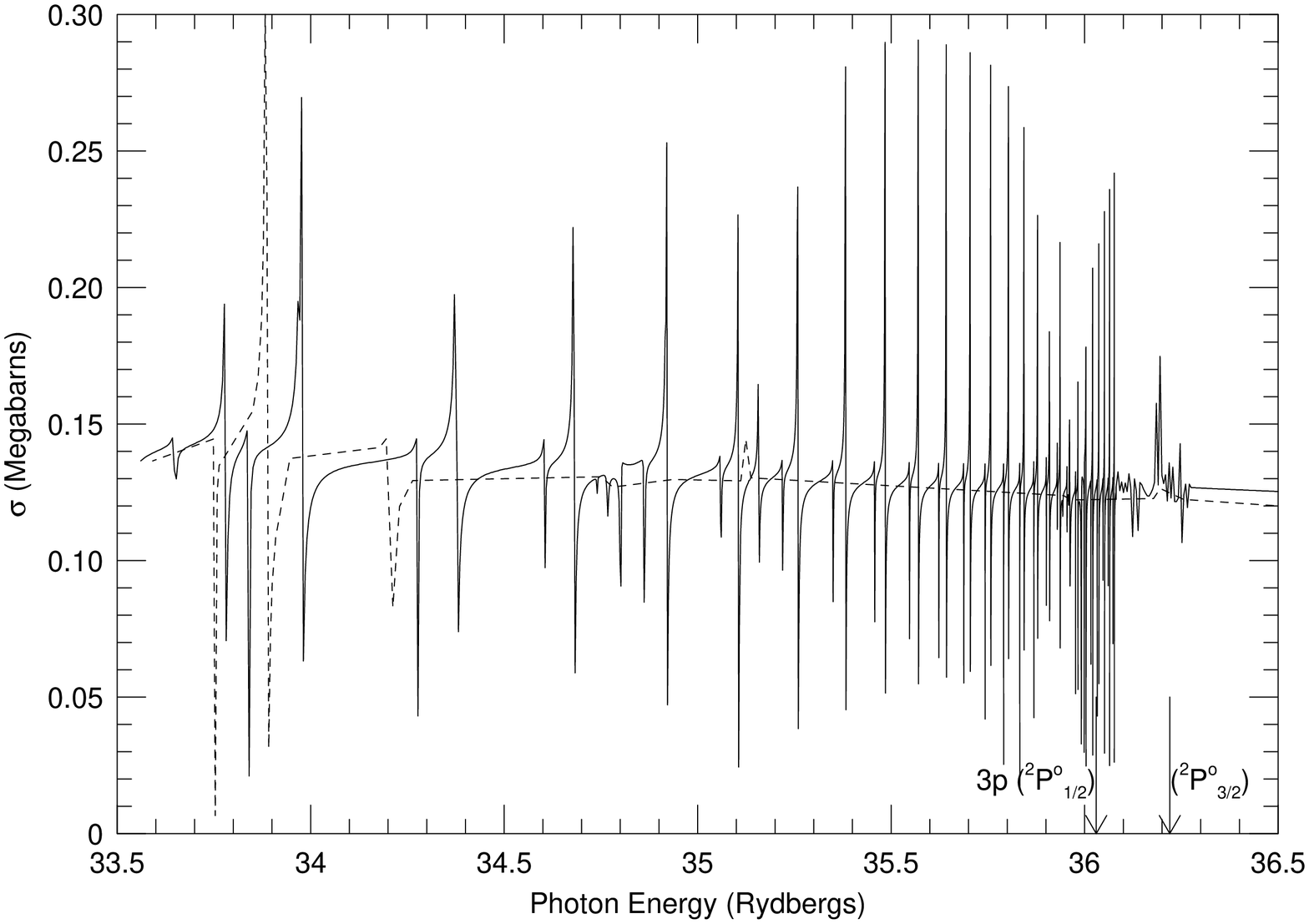,height=10.0cm,width=10.0cm}
\caption{Photoionization cross section of the ground level of Fe~XV
$3s^2 \
(^1S_0)$ in the near-threshold region; solid line - relativistic BPRM
results, dashed line - non-relativistic results [OP,8].}
\end{figure}

\begin{figure}
\centering
\psfig{figure=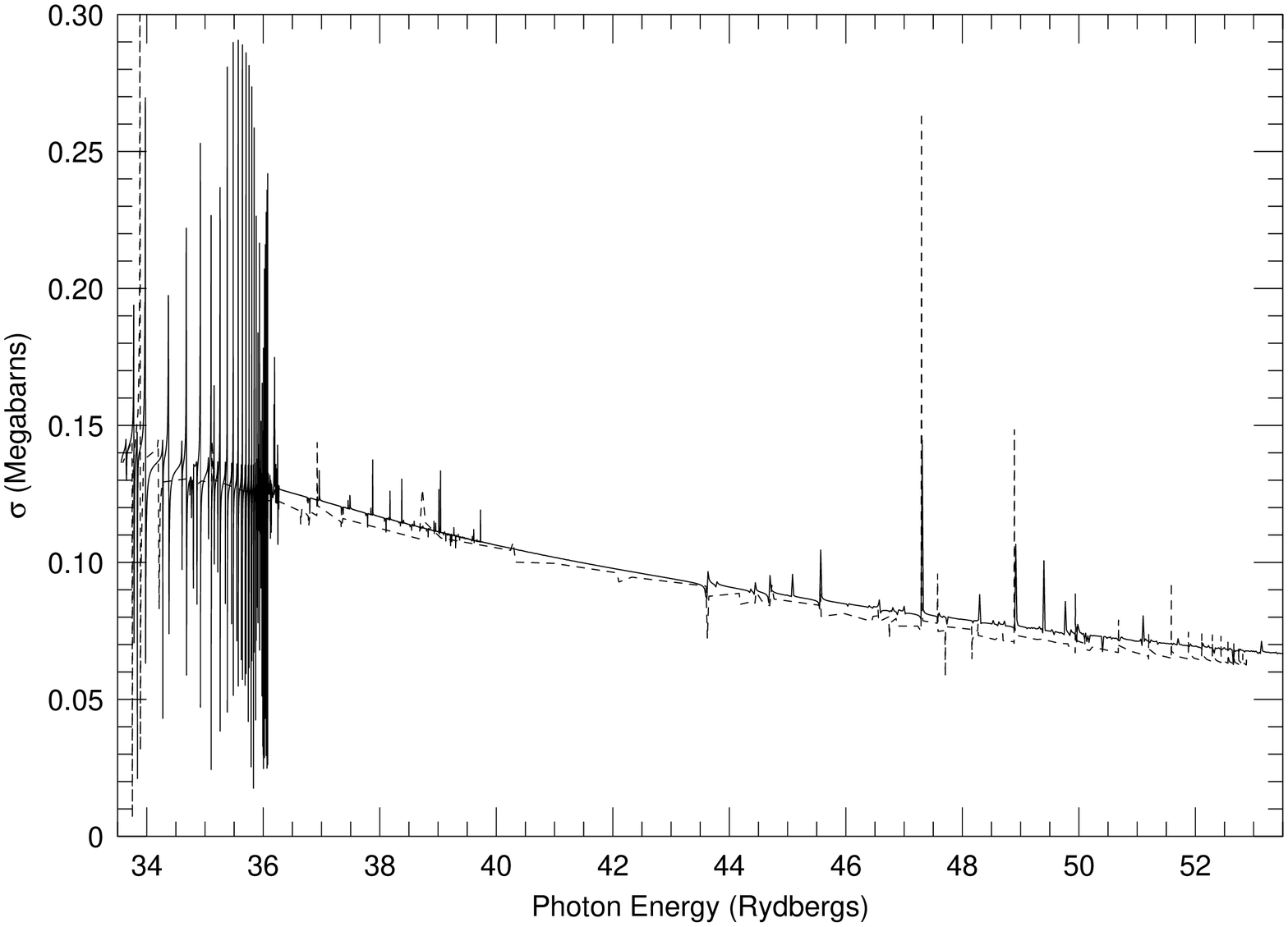,height=10.0cm,width=10.0cm}
\caption{Comparison of the present BPRM cross sections with the earlier
OP non-relativistic calculations over an extended range spanning all
target
thresholds in the close coupling expansion.}
\end{figure}

 BPRM calculations can be further extended in the intermediate energy
range, from the highest target level to the n = 2 inner-shell 
ionization thresholds. This region does not include the target levels 
that would correspond to the first sum in the close coupling expansion 
(Eq. [1]). However, a number of (N+1)-electron configurations, ${\Phi_j(Fe~XV)}$, 
are included in the BPRM calculations in the second sum in Eq. [1]) with
eigenvalues in this range. These
Fe~XV configurations correspond to outer 3s electron excitation, as
well as the inner 2p electron excitations, i.e. $[1s^22s^22p^6] \ 
3 \ell^2 + 3 \ell, n \ell; n \leq 5, \ell \leq 2$, and the inner-shell
excitation configurations: $[1s^22s^22p^5] \ 3\ell^2 n\ell; n \leq 5,
\ell \leq 2$. Although the 2p-shell ionization are not
explicitly represented in the BPRM calculations, the 2p-shell
excitation-autoionization is included to some extent 
via resonances corresponding to
these ${\Phi_j(Fe~XV)}$. Some of the resonances in the 30 - 80 Ry region,
 due to inner-shell photoexcitations, are much stronger than those 
due to the valence 3s-shell excitations in Figs. 1 and 2, 
as discussed below (Fig. 4). 

 Close to the ionization threshold of the n = 2, we carry out RRPA+RMQDT
calculations, with the detailed resonance strucutures shown
in Fig. 3.
The RRPA and RMQDT calculations are carried out as described in [9]
and [10]. In the autoionization resonance spectrum region between 80-86
Ry, the photoionization cross section in open dissociation channels,
that is, the 3s photoionization cross section, is calculated and plotted
as a function of photon energy in Fig.3. The seven interacting
relativistic dipole channels considered for the RRPA+RMQDT calculations
are: $3s \longrightarrow \epsilon (p_{3/2}) , \epsilon(p_{1/2}) ; 2p3/2 
\longrightarrow nd_{5/2,3/2}, ns_{1/2}$; and $2p_{1/2} \longrightarrow
nd_{3/2}, ns_{1/2}$. 
The theoretical single configuration Dirac-Fock thresholds for $3s,
2p_{3/2}, 2p_{1/2}$  are at 33.314, 86.26, and 87.21 Ry respectively.
Below the $2p_{3/2}$
threshold there are five interacting Rydberg series corresponding to
discrete excitation from $2p_{3/2}$ and $2p_{1/2}$ levels. 
Coupling between these
excitation channels and the ionization channels from the 3s levels causes
configuration interaction between the discrete and continuum states,
which in turn leads to autoionization resonances seen in Fig.3.

\begin{figure}
\centering
\psfig{figure=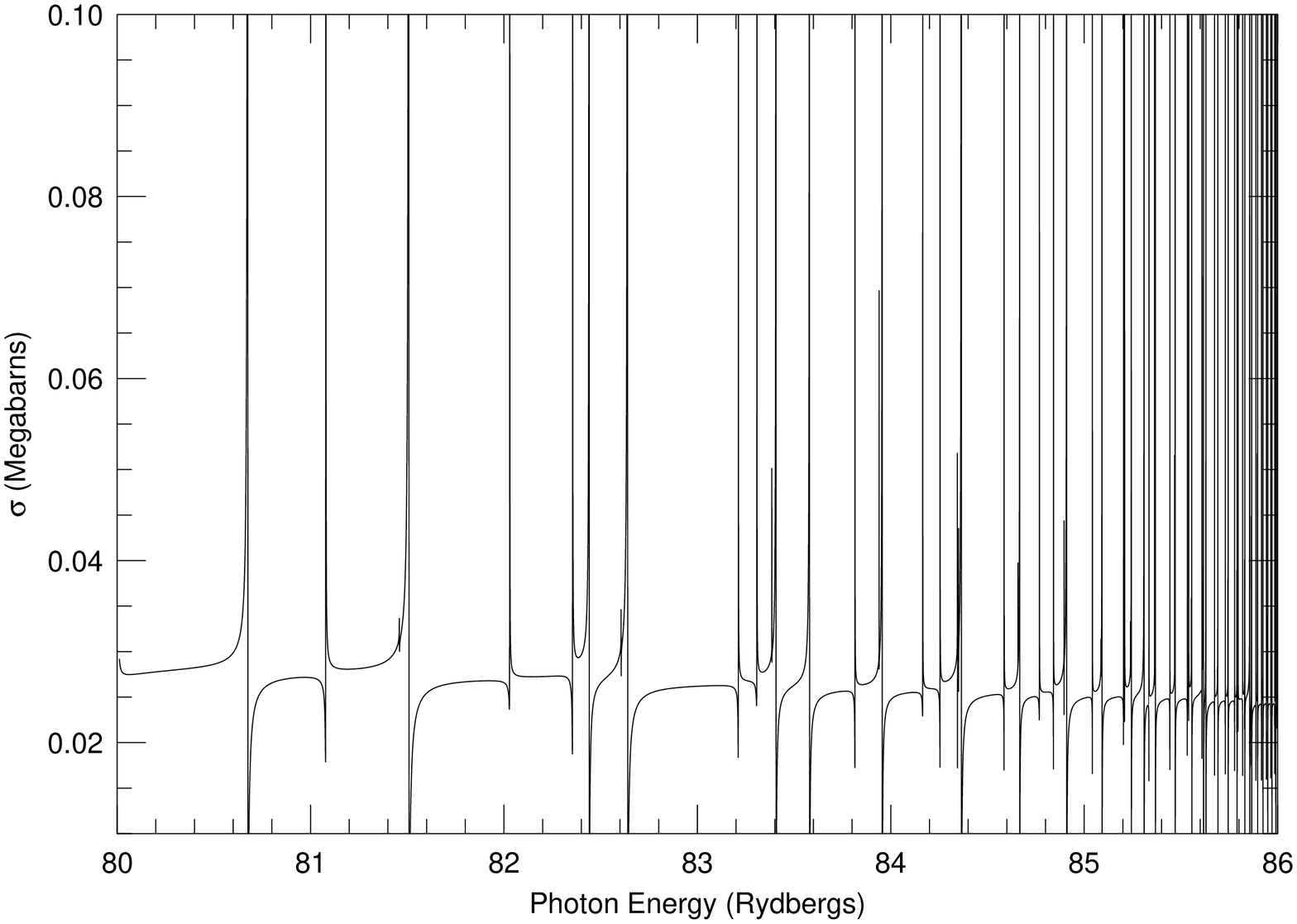,height=10.0cm,width=10.0cm}
\caption{RRPA and RMQDT cross sections
with resonances converging on to the L-shell ionization thresholds.}
\end{figure}

 Finally, Fig. 4 shows the complete set of BPRM and RMQDT+RRPA  results
obtained. The strongest resonances in the middle range between 54 - 70
Ry correspond mainly to inner-shell excitation configurations 
$2p^5 \ 3s^2 \ (3p , 3d) $,  particularly to the $2p \longrightarrow 3d$
excitation. We identify these resonances from
an extensive atomic structure configuration-interaction 
calculation for Fe~XV using SUPERSTRUCTURE,
employing the same one-electron orbital basis set as the target ion Fe~XVI.
For example, the autoionizing levels $2p^5 \ 3s^2 \ 3p \ 
^{3,1}(S_1,P_{0,1,2},D_{0,1,2,3})$, 
lie between 54.49 - 56.83 Ry. The large resonance complex 
between 58.26 - 59.76 Ry corresponds to the autoionizing levels 
$2p^5 \ 3s^2 \ 3d \ ^{3,1}(S_1,P_{0,1,2},D_{0,1,2,3})$. The experimental
first ionization potential of Fe~XV is 33.58 Ry (the present BPRM
calculated value is 33.56 Ry), and the highest Fe~XVI target
threshold is at 25.41 Ry. This implies that the 
L-shell resonance complex just below 60 Ry coincides 
with the highest target levels $5d(^2D_{3/2,5/2})$ in the 
wavefunction expansion (Table 1). It is clear that
the inner-shell photoexcitation-autoionization would be a major
contributor to the effective photoionization cross sections {\it below} the
L-shell ionization threshold(s).

\begin{table}
\begin{minipage}{80mm}
\caption{Fe~XVI target level energies (Ry) in the wavefunction expansion}
\begin{center}
\begin{tabular} {lllll}
\hline
 Index & Configuration [Term] & J & E(Obs) & E(Calc) \\
\hline
  1 & $2p^6(^1S)3s \ [^2S]   $  &   1/2 &  0.00000 & 0.00000           \\   
  2 & $2p^6(^1S)3p \ [^2P^o] $  &   1/2 &  2.52596 & 2.52186           \\
  3 & $2p^6(^1S)3p \  [^2P^o] $  &   3/2 &  2.71688 & 2.69095           \\
  4 & $2p^6(^1S)3d \ [^2D]   $  &   3/2 &  6.15544 & 6.21637           \\
  5 & $2p^6(^1S)3d \  [^2D]   $  &   5/2 &  6.18198 & 6.25459           \\
  6 & $2p^6(^1S)4s \ [^2S]   $  &   1/2 &  17.0182 & 17.0836           \\
  7 & $2p^6(^1S)4p \ [^2P^o] $  &   1/2 &  18.0252 & 18.0643           \\
  8 & $2p^6(^1S)4p \ [^2P^o] $  &   3/2 &  18.0980 & 18.1269           \\
  9 & $2p^6(^1S)4d \ [^2D]   $  &   3/2 &  19.3570 & 19.4013           \\
 10 & $2p^6(^1S)4d \ [^2D]   $  &   5/2 &  19.3677 & 19.4186           \\
 11 & $2p^6(^1S)4f \ [^2F^o] $  &   5/2 &  19.9077 & 19.9638           \\
 12 & $2p^6(^1S)4f \ [^2F^o] $  &   7/2 &  19.9125 & 19.9703           \\
 13 & $2p^6(^1S)5s \ [^2S]   $  &   1/2 &  24.2500 & 24.3164           \\
 14 & $2p^6(^1S)5p \ [^2P^o] $  &   1/2 &  24.7606 & 24.8020           \\
 15 & $2p^6(^1S)5p \ [^2P^o] $  &   3/2 &  24.7970 & 24.8357           \\
 16 & $2p^6(^1S)5d \ [^2D]   $  &   3/2 &  25.4065 & 25.4552           \\
 17 & $2p^6(^1S)4d \ [^2D]   $  &   5/2 &  25.4116 & 25.4639           \\
\hline

\end{tabular}
\end{center}
\end{minipage}
\end{table}

\begin{figure}
\centering
\psfig{figure=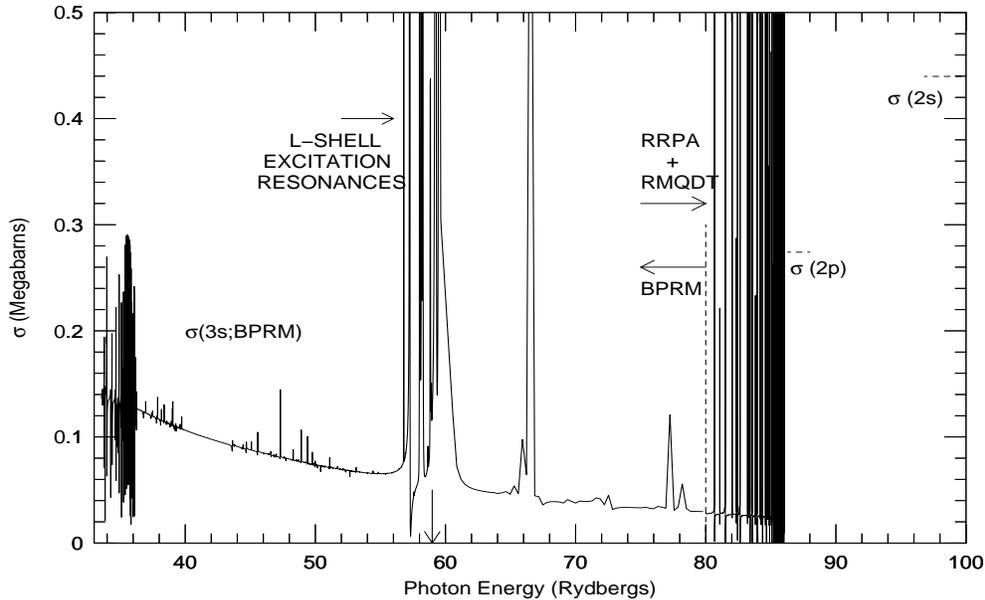,height=10.0cm,width=15.0cm}
\caption{The BPRM and (RRPA + RMQDT) cross sections in the energy range up to
L-shell ionization. The arrow at 59.98 Ry is the highest target
threshold in the BPRM calculations. The total sub-shell photoionization
cross sections $\sigma(2p)$ and $\sigma(2s)$
are computed using the RRPA.}
\end{figure}

 The RRPA and RMQDT results in Figs. 3 and 4 show an excellent match with 
the background BPRM cross sections. The resonances just below the 2p
ionization threshold form a pseudocontinuum whose effective area
corresponds to the total photoabsorption in this region, which in turn
rises to the threshold value at the 2p ionization energy of
86.26 Ry. The total photoabsorption cross section should be
continuous across ionization thresholds. Therefore the cross section at
and above threshold(s) is, in principle, comparable in magnitude 
to the effective resonance averaged
photoexcitaion-autoionization cross section below threshold(s). 
The energy region up to the 2p and 2s ionization thresholds 
in Fig. 4 is dominated by resonance structures, many of which are not yet
considered. The inner-shell ionization edges in particular are not
sharp jumps, as often obtained in non-resonant calculations, but
as diffuse pseudocontinua rising up to the relevant threshold cross
sections.

As these first
detailed calculations including relativistic and resonance effects 
show, the BPRM close coupling calculations with
the large number of
levels and channels in the intermediate energy range should considerably
enhance the photoionization cross section of Fe~XV. Thus, calculations
that neglect the extensive inner-shell excitation resonances are likely
to underestimate the cross sections by large factors (as has been shown, for
example, in the case of photoionization of neutral iron [10]). 
Although likely to be much more extensive, the present work emphasizes
the urgent need for more elaborate BPRM calculations  leading
up to the L- and K-shell ionization thresholds, as these data would be
essential requirements for photoionization modeling of astrophysical
X-ray sources observed by CXO, XMM, and Astro-E.

\acknowledgments

This work was partially supported by the National Science Foundation
 and the NASA Astrophysical Theory Program. N. Haque would like to thank
Prof. Walter Johnson for the use of his RRPA and RMQDT codes.

%***
%***  E n d   o f   p a g e   1   o f   g a l l e y - m o d e   o u t p u t
%***

\narrowtext

\end{document}